\documentclass[conference]{IEEEtran}

\ifCLASSINFOpdf
\else
\fi
\usepackage{amsmath}
\usepackage{amsfonts}
\usepackage{algorithm}
\usepackage{algorithmic}

\usepackage{array}

\usepackage{url}

\usepackage[]{fancyhdr} %
\newcommand{\changefont}{\fontsize{9}{9}\selectfont}
\usepackage{caption}
\usepackage{subcaption}

\usepackage{tablefootnote}
\usepackage{braket}
\usepackage{bm}
\usepackage{xcolor}
\usepackage{circuitikz}

\usetikzlibrary{shapes.geometric,arrows,calc}
\tikzstyle{platform} = [rectangle, rounded corners, minimum width=10mm, minimum height=10mm,text centered, draw=black]
\tikzstyle{lab} = [rectangle, minimum width=5mm, minimum height=2mm, text centered, draw=black,fill=white]
\tikzstyle{method} = [rectangle, minimum width=5mm, minimum height=5mm, text centered, draw=black,fill=gray!30]
\tikzstyle{arrow} = [thick,->,>=stealth]

\tikzset{
    load/.style={
        draw,
        shape border rotate=180,
        regular polygon,
        regular polygon sides=3,
        fill=black,
        node distance=0.1cm,
        minimum height=0.1cm
    }
}

\makeatletter
\newcommand\notsotiny{\@setfontsize\notsotiny\@vipt\@viipt}
\makeatother

\IEEEoverridecommandlockouts
\begin{document}

%

\title{Quantum Computing for Power Flow Algorithms: Testing on real Quantum Computers}

\author{\IEEEauthorblockN{Brynjar Sævarsson, Spyros Chatzivasileiadis, Hjörtur Jóhannsson, Jacob Østergaard}
\IEEEauthorblockA{Department of Wind and Energy Systems\\Division of Electric Power and Energy\\Technical University of Denmark\\
Kgs. Lyngby, Denmark \\ \{brysa, spchatz, hjjo, jaos\}@dtu.dk}} 

\maketitle
\thispagestyle{fancy}
\pagestyle{fancy}

\begin{abstract}
Quantum computing has the potential to solve many computational problems exponentially faster than classical computers. 
The high shares of renewables and the wide deployment of converter-interfaced resources require new tools that shall drastically accelerate power system computations, including optimization and security assessment, which can benefit from quantum computing. 
To the best of our knowledge, this is the first paper that goes beyond quantum computing simulations and performs an experimental application of Quantum Computing for power systems on a real quantum computer.
We use five different quantum computers, apply the HHL quantum algorithm, and examine the impact of current noisy quantum hardware on the accuracy and speed of an AC power flow algorithm. We perform the same studies on a 3-bus and a 5-bus system with real quantum computers to identify challenges and open research questions related with the scalability of these algorithms.

\end{abstract}
 



%
\IEEEpeerreviewmaketitle

\section{Introduction}

The increasing penetration of distributed renewable energy sources (RES) brings a number of challenges when it comes to ensuring a secure operation of power systems. With traditional synchronous generation being replaced by thousands of converter-interfaced generation sources, the complexity of the power system greatly increases, its natural inertia decreases, and the dynamics of the system change. This means traditional offline methods for security assessment are expected to become insufficient as more detailed, and computationally demanding, simulations are required to capture the faster dynamics of RES. Additionally, the operating state of the system becomes subject to the prevailing weather conditions, making it difficult to predict hours ahead of time. The uncertainties introduced by RES can lead to exponentially more contingency scenarios needed to be considered to ensure N-x security. These challenges drive the need for new real-time and offline tools for security analysis. Assessing large complex systems for millions of contingency scenarios can be very computationally demanding and is currently one of the major challenges that utilities are expected to face in the future. 

Over the last few years there has been a great leap in the area of Quantum Computing (QC), bringing us into the so called Noisy Intermediate-Scale Quantum (NISQ) era \cite{nisq} of quantum computing, where real quantum computers are already containing over 100 qubits. According to roadmaps of vendors such as IBM \cite{roadmap}, there is ambitious development in building scalable quantum computers, with the aim of creating large-scale and noise-free devices in the near future. However, the current technology readiness level (TRL) of the NISQ-era quantum computers is still rather low and only very basic applications can be implemented. The use of quantum computations for power system applications is very new. Quantum power flow algorithms have been proposed to solve DC power flow \cite{q_dc_pf} and AC power flow \cite{q_ac_pf}, and were tested with simulated quantum computers. The use of QC for contingency assessment was introduced in \cite{q_sec} and for performing EMT simulations in \cite{q_emt}. A common feature of these papers is the use of the HHL quantum algorithm for solving linear equations \cite{hhl}, as this promises exponential speedup compared to classical methods.

To the best of our knowledge, this paper presents the first implementation of an AC power flow on real Quantum Computers. 
The intention with this paper is to explore the current capabilities of QC for power flow studies, which challenges we need to address, and to investigate the foreseen future capabilities and practical implications of QC for power systems.
We test our power flow algorithm for a simple 3 bus system on four of IBM's publicly available quantum computers to investigate the impact of noisy hardware. For a 5 bus system we use one of IBM's larger quantum computers to investigate how the method scales with increasing system size. This is important for future implementations of full-scale power systems since the benefits of the computational power of quantum computing will be more apparent for large power systems. Larger systems exceed the current hardware capabilities of real quantum computers and are not tested in this paper.

This paper is organized as follows. Section~\ref{sec:qc} provides an overview of quantum computing and its potential for power systems. Section~\ref{sec:pf} describes the quantum power flow method and its implementation. In Section~\ref{sec:sim}, we describe the simulation setup and the quantum hardware requirements. Section~\ref{sec:results} provides the results of the power flow on both real and simulated quantum computers. In Section~\ref{sec:discussion}, we discuss potential issues with the scalability of the method. Section~\ref{sec:conclusion} concludes.

\section{The potential of quantum computing for power systems}
\label{sec:qc}
The way quantum computers function is fundamentally different from classical computers. Instead of using classical bits, which can only be in the states 0 and 1, quantum computers use quantum bits (qubits) which, in addition to being in states 0 and 1, can form a linear combination of states, i.e. have a certain probability to be measured either as 0 or 1. This state is referred to as superposition. While in superposition, they can also be said to be in all states simultaneously, i.e. both 0 and 1 or somewhere in-between, and only when measured do they collapse to 1 or 0 with some probability. An important feature of this is that the information contained in a quantum register grows exponentially for every qubit which is added \cite{qcbook}. That is, one could say that as a rule of thumb, instead of requiring $2^n$ bits for a computing process with classical computing, quantum computing requires only $n$ qubits. For example, if a process would require 1024 bits in a classical computer, a quantum computer would aprroximately only need 10 qubits. Please bear in mind that currently publicly available quantum computers do not exceed 7 qubits. 
Currently existing Quantum Computers (but not publicly available), on the other hand, go up to 100 qubits. 
That is a 30 orders of magnitude more powerful computer, while we have increased the number of qubits by only 1 order of magnitude.

With qubits being able to be in superposition, i.e. 0 or 1 or somewhere in-between, quantum states can also be used to evaluate functions for multiple values simultaneously. In a power systems context, such a feature can become extremely valuable for power flow analysis of multiple scenarios. Instead of sequentially solving multiple power flows to assess a large number of generation and demand scenarios, theory suggests that future QC infrastructure can deliver the solution for many scenarios simultaneously, drastically accelerating e.g. the security assessment. 
This, and other special features of QC, allows them to solve certain problems exponentially faster than classical computers. Theoretically, Quantum Computers can solve all the same problems as classical computers; not all problems, though, allow for exponential speedups and in some cases QC are even slower than classical computers. QC is therefore seen as a supplement to classical computing for solving specific problems in a hybrid quantum-classical computation.

In the next sections, we present the formulation of the quantum power flow algorithm and the results we obtain after testing it on real quantum computers. Considering that the quantum computers that were accessible at the time of this paper only have 7 qubits, the application of our quantum power flow algorithm on real quantum computers is limited to a 5-bus system.

\section{Quantum Power Flow method}
\label{sec:pf}
The implementation of the quantum power flow method we follow in this paper is based on the Fast Decoupled Load Flow (FDLF) method, which is a commonly used adaptation of the Newton Raphson Power Flow (NRPF). It exploits inter-dependence between $P-\theta$ and $Q-|V|$ to create two constant Jacobian sub-matrices $B^{'}$ and $B^{''}$. This replaces one of the most computationally heavy parts of NRPF i.e. updating the Jacobian matrix in every iteration. FDLF converges to the same solution as NRPF since the mismatch functions are the same, but it usually requires more iterations than NRPF \cite{fdlf}. On the other hand, each iteration of FDLF is much faster, meaning that the overall computation time of FDLF can often be quite faster, especially for large systems. In each iteration, we solve equations \eqref{eq_mis}-\eqref{eq_dq} for $\Delta \theta$  and $\Delta V_{m}$ until the norm of $\Delta P$  and $\Delta Q$ is less than the chosen tolerance $\xi$.

\begin{IEEEeqnarray}{lCr}
\Delta S = (S_{bus} - \bm{V} \circ \overline{(Ybus \cdot \bm{V})})\oslash V_{m} \label{eq_mis} \\
\Delta P = \Re(\Delta S_{pv+pq})=B^{'}\Delta \theta \label{eq_dp} \\
\Delta Q = \Im(\Delta S_{pq})=B^{''}\Delta V_{m} \label{eq_dq} 
\end{IEEEeqnarray}

In \eqref{eq_mis}-\eqref{eq_dq}, $\bm{V}=V_{m}\angle\theta$ is a vector collecting the voltage phasors for all buses, $\circ$ and $\oslash$ denote the element-wise product and division respectively, and $pv$ and $pq$ are indices corresponding to the set of PV and PQ buses respectively. The structure of FDLF makes it very suitable for hybrid classical-quantum computing since we can easily replace the numerical algorithm solving equations \eqref{eq_dp} and \eqref{eq_dq} with a method which is capable of running on a quantum computer.
HHL is probably the most popular quantum algorithm for solving a set of linear equations at the moment, with a runtime of $\mathcal{O}(log(N)\frac{s^{2}\kappa^{2}}{\epsilon})$ \cite{hhl}; here, N is the number of equations, $s$ is the sparsity and $\kappa$ the condition number of the system matrix, and $\epsilon$ is the solution accuracy. In theory, HHL can achieve an exponential speedup over the fastest classical algorithm, the conjugate gradient method, which has a runtime of $\mathcal{O}(Ns\kappa log(\frac{1}{\epsilon}))$. 

A system of linear equations is usually given in the form:
\begin{equation}
    Ax=b \label{eq_sle},
\end{equation}
where, given a known matrix $A$ and vector $b$, we solve for $x$. For quantum systems we use the Dirac (or bra-ket) notation, where we denote a vector with a \textbf{ket} which has the form $\bm{\ket{v}}$ and represents the state of a quantum system. We also denote a \textbf{bra} with the form $\bm{\bra{v}}$, where bra is the conjugate transpose of ket. Quantum vectors are often very sparse, i.e. with most amplitudes equal to zero, and this notation makes it much simpler to represent values of interest. A single qubit system can be described as a combination of the two states $\ket{0}$ and $\ket{1}$:
\begin{equation}
    \ket{q}=\alpha\ket{0}+\beta\ket{1},
    \label{eq: qubit}
\end{equation}
where the probability of it being measured as 0 is $|\alpha|^2$ and the probability of it being measured as 1 is $|\beta|^2$. The combined probability amplitudes must satisfy that $|\alpha|^2+|\beta|^2=1$. As also mentioned earlier, this is one of the key strengths of quantum computing: as qubits can be in both states at the same time, they can represent multiple variables. For a single qubit, for example, we can associate one variable with state 0 and one variable with state 1. 
The probability of the qubit to be measured as 1 corresponds to the amplitude of that variable. Let us assume we encode $\Delta \theta_1$ to the qubit state $\ket{0}$ and $\Delta \theta_2$ to the qubit state $\ket{1}$. The value of $\Delta \theta_1$ is then equal to $C \cdot \alpha$, where $C$ is a scaling factor. As we will see in Section~\ref{sec:results} and \eqref{eq:qubit_prob}, the scaling factor $C$ depends on how we have normalized our input to be between 0 and 1. 


Going back to the solution of a linear system with the HHL algorithm, we rescale the system in \eqref{eq_sle} by normalizing $x$ and $b$, and we can then map them to the quantum states $\ket{x}$ and $\ket{b}$. 

\begin{equation}
    A\ket{x}=\ket{b} \label{eq_qsle}
\end{equation}

Portraying equation \eqref{eq_dp} in this form we get:

\begin{equation}
    B^{'}\ket{\Delta \theta}=\ket{\Delta P} \label{eq_qdp}
\end{equation}

If we express the right hand side in the eigenbasis of the $B^{'}$ matrix, we can write it as: 
\begin{IEEEeqnarray}{lCr}
    \ket{\Delta P} =\sum_{j=0}^{N-1} \Delta P_{j} \ket{u^{'}_{j}} \label{eq_ketdp} 
\end{IEEEeqnarray}

The HHL method requires matrix $A$ to be Hermitian, that is equal to its conjugate transpose. For non-hermitian matrices, a hermitian matrix can be constructed from A such that:

\begin{equation}
\begin{bmatrix}
0 & A \\
A^{\dagger} & 0
\end{bmatrix}
\begin{bmatrix}
0 \\
x \end{bmatrix} =
\begin{bmatrix}
b \\
0
\end{bmatrix}
\end{equation}
However, $B^{'}$ and $B^{''}$ of FDLF happen to be hermitian. And since the matrices are Hermitian, they have a spectral decomposition:
\begin{IEEEeqnarray}{lCr}
    B^{'} =\sum_{j=0}^{N-1} \lambda^{'}_{j} \ket{u^{'}_{j}}\bra{u^{'}_{j}} \\
    B^{'-1} =\sum_{j=0}^{N-1} \lambda^{'-1}_{j} \ket{u^{'}_{j}}\bra{u^{'}_{j}} \label{eq_binv}
\end{IEEEeqnarray}

Where $ \lambda_{j}^{'} $ and $ u_{j}^{'}$ are the $j_{th}$ eigenvalue and eigenvector of the $B^{'}$ matrix. Putting \eqref{eq_qdp}, \eqref{eq_ketdp} and \eqref{eq_binv} together we then get:

\begin{IEEEeqnarray}{lCr}
    \ket{\Delta \theta}=B^{'-1}\ket{\Delta P}=\sum_{j=0}^{N-1} \lambda^{'-1}_{j} \Delta P_{j} \ket{u^{'}_{j}}\bra{u^{'}_{j}} \ket{u^{'}_{j}} \label{eq_dth1} 
\end{IEEEeqnarray}

As the scalar product $\bra{u^{'}_{j}} \ket{u^{'}_{j}}=1 $, \eqref{eq_dth1} simplifies to:
\begin{IEEEeqnarray}{lCr}
    \ket{\Delta \theta}=\sum_{j=0}^{N-1} \lambda^{'-1}_{j} \Delta P_{j} \ket{u^{'}_{j}} \label{eq_dth2} 
\end{IEEEeqnarray}
And following the same procedure for $\Delta V_{m}$ and $B^{''}$ we get:
\begin{IEEEeqnarray}{lCr}
    \ket{\Delta V_{m}}=B^{''-1}\ket{\Delta Q}=\sum_{j=0}^{N-1} \lambda^{''-1 }_{j} \Delta Q_{j} \ket{u^{''}_{j}} \label{eq_dv}
\end{IEEEeqnarray}

\begin{figure}[h!]
\centering
\includegraphics[width=0.95\columnwidth]{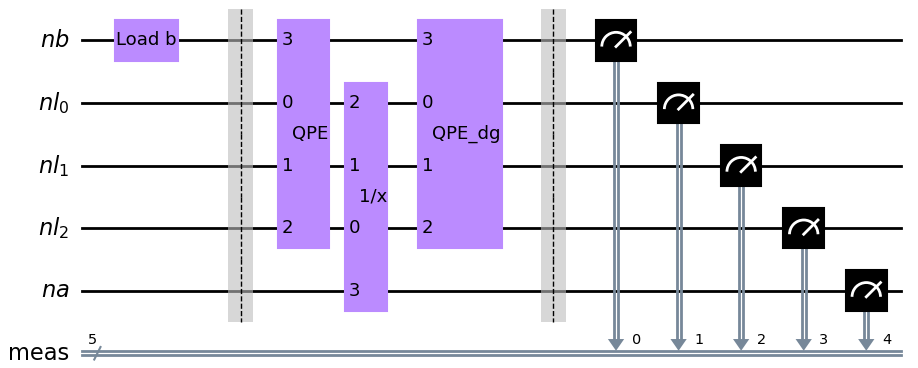}
\caption{Quantum HHL circuit used for solving a Quantum Power Flow (QPF) on a 3-bus system. The blocks shown in the circuit are: Preparation of the data vector (Load b), Quantum Phase Estimation (QPE), conditioned rotation (1/x) and inverse QPE (QPE\_dg). Finally, each qubit is measured and the result is stored in a classical 5 bit register where ``meas 0'' is the least significant bit and ``meas 4'' is the most significant bit.}
\label{fig_circuit}
\end{figure}

 To solve \eqref{eq_dth2} a quantum circuit is implemented as shown in Fig.~\ref{fig_circuit}. This circuit is for solving a 2x2 system of equations, i.e., $N=2$. It consists of three registers, $nb,nl,na$, which are all initialized as $\ket{0}$. $nb$ is the data register where we load the data vector $b$ (here $b \equiv \Delta P$ onto $log_{2}(N)=1$ qubits):

\begin{equation}
    \ket{0} \mapsto \ket{\Delta P}
\end{equation}

Next, we apply quantum phase estimation to estimate the eigenvalues of the $B$ matrix. The $nl$ register consists of 3 qubits where we store the approximation of the eigenvalues $\ket{\Tilde{\lambda_j}}$. The more qubits used in $nl$, the more accurately we can estimate the eigenvalues of $B$. The last register, $na$, consists of a single auxiliary qubit. The output auxiliary qubit indicates if we can obtain a binary estimation of the eigenvalues and the data vector, i.e. if we can trust their numerical result when we measure them, or not. To get the output of the auxiliary qubit, we perform a rotation conditioned on $\ket{\Tilde{\lambda_j}}$, and finally an inverse phase estimation. When we measure the output of the auxiliary qubit as 1, then the registers $nb, nl$ are in the post measurement state; this means that the phase estimation outputs a binary estimation of the eigenvalues of the B matrix, and after an inverse phase estimation, the $nb$ register contains the solution $\ket{\Delta \theta}$. If the auxiliary qubit is 0, then the states are considered to contain no useful information and are, therefore, discarded. For more information about the steps of the HHL algorithm, the interested reader can refer to \cite{hhl}.

The implemented Quantum Power Flow (QPF) algorithm is shown in Algorithm \ref{alg_qpf}. 

\begin{algorithm}
\caption{Quantum Power Flow}
\begin{algorithmic}
\STATE \textbf{Input} $Y_{bus}, B^{'}, B^{''}, V_0, S_{bus}$ 
\STATE Update $\Delta P$, $\Delta Q$
\WHILE{ $\lVert \Delta P \rVert \geq \xi$ \textbf{AND} $\lVert \Delta Q \rVert \geq \xi$}
\STATE $\Delta \theta = HHL(B^{'}, \Delta P) $
\STATE $\Delta V_{m} = HHL(B^{''}, \Delta Q) $
\STATE Update $V_{m}\angle\theta$
\STATE Update $\Delta P$, $\Delta Q$
\ENDWHILE
\RETURN $V_{m}\angle\theta$
\end{algorithmic}
\label{alg_qpf}
\end{algorithm}

\section{Simulations}
\label{sec:sim}

\begin{figure}
    \centering
    \begin{subfigure}[b]{0.5\columnwidth}
        \def\rowa{0.0cm}
\def\rowb{-1.0cm}
\def\cola{0.0cm}
\def\colb{0.75cm}
\def\colc{1.5cm}

\def\gs{0.7cm} 
\def\dl{0.2cm} 
\def\dg{0.7cm} 
\def\wg{0.4cm} 
\def\bw{0.5cm} 
\def\bt{0.1cm} 
\def\lt{0.01cm} 
\def\lo{0.3cm} 
\def\ls{0.35} 
\def\dld{0.3cm} 

\begin{center}
\begin{circuitikz}[scale=1]

\draw [line width=\bt] (\cola-\bw,\rowa) -- (\cola+\bw,\rowa);
\draw [line width=\bt] (\colb-\bw,\rowb) -- (\colb+\bw,\rowb);
\draw [line width=\bt] (\colc-\bw,\rowa) -- (\colc+\bw,\rowa);

\draw(\cola+\lo,\rowa) node[above=-0.05cm] {\footnotesize{1}};
\draw(\colb+\lo,\rowb) node[below=-0.05cm] {\footnotesize{3}};
\draw(\colc+\lo,\rowa) node[above=-0.05cm] {\footnotesize{2}};

\draw (\cola,\rowa+\dg) to [/tikz/circuitikz/bipoles/length=\gs,sV,l_=\tiny$G_1$] (\cola,\rowa+\dg-\wg);
\draw (\cola,\rowa+\dg-\wg) to[short] (\cola,\rowa);
\draw (\colc,\rowa+\dg) to [/tikz/circuitikz/bipoles/length=\gs,sV,l_=\tiny$G_2$] (\colc,\rowa+\dg-\wg);
\draw (\colc,\rowa+\dg-\wg) to[short] (\colc,\rowa);

\draw [line width=\lt] (\cola+\lo,\rowa) -- (\cola+\lo,\rowa-\dl ) -| (\colc-\lo,\rowa);
\draw [line width=\lt] (\cola-\lo,\rowa) -- (\cola-\lo,\rowa-\dl ) -- (\colb-\lo,\rowb+\dl ) -- (\colb-\lo,\rowb);
\draw [line width=\lt] (\colc+\lo,\rowa) -- (\colc+\lo,\rowa-\dl ) -- (\colb+\lo,\rowb+\dl ) -- (\colb+\lo,\rowb);

\node [load,shape border rotate=180,scale=\ls] at (\colb,\rowb-\dld) {};
\draw (\colb,\rowb) to[short] (\colb,\rowb-\dld);

\end{circuitikz}
\end{center}

\vspace{-5mm}
        \caption{Three bus system} 
        \label{fig_five_bus}
    \end{subfigure}%
    ~ 
    \begin{subfigure}[b]{0.5\columnwidth}
        \def\rowa{0.0cm}
\def\rowb{-1.0cm}
\def\cola{0.0cm}
\def\colb{1.25cm}
\def\colc{2.50cm}

\def\gs{0.7cm} 
\def\dl{0.2cm} 
\def\dg{0.7cm} 
\def\wg{0.4cm} 
\def\bw{0.5cm} 
\def\bt{0.1cm} 
\def\lt{0.01cm} 
\def\lo{0.3cm} 
\def\ls{0.35} 
\def\dld{0.3cm} 

\begin{center}
\begin{circuitikz}[scale=1]

\draw [line width=\bt] (\cola-\bw,\rowa) -- (\cola+\bw,\rowa);
\draw [line width=\bt] (\colb-\bw,\rowa) -- (\colb+\bw,\rowa);
\draw [line width=\bt] (\colc-\bw,\rowa) -- (\colc+\bw,\rowa);
\draw [line width=\bt] (\cola-\bw,\rowb) -- (\cola+\bw,\rowb);
\draw [line width=\bt] (\colc-\bw,\rowb) -- (\colc+\bw,\rowb);

\draw(\cola+\lo,\rowa) node[above=-0.05cm] {\footnotesize{2}};
\draw(\cola-\lo-0.15cm,\rowb) node[above=-0.05cm] {\footnotesize{1}};
\draw(\colb+\lo,\rowa) node[above=-0.05cm] {\footnotesize{3}};
\draw(\colc+\lo,\rowa) node[above=-0.05cm] {\footnotesize{4}};
\draw(\colc-\lo,\rowb) node[above=-0.05cm] {\footnotesize{5}};
 
\draw (\cola,\rowa+\dg) to [/tikz/circuitikz/bipoles/length=\gs,sV,l_=\tiny$G_2$] (\cola,\rowa+\dg-\wg);
\draw (\cola,\rowa+\dg-\wg) to[short] (\cola,\rowa);
\draw (\cola,\rowb-\dg) to [/tikz/circuitikz/bipoles/length=\gs,sV,l_=\tiny$G_1$] (\cola,\rowb-\dg+\wg);
\draw (\cola,\rowb-\dg+\wg) to[short] (\cola,\rowb);

\draw [line width=\lt] (\cola-\lo,\rowa) -- (\cola-\lo,\rowb);
\draw [line width=\lt] (\cola+\lo,\rowa) -- (\cola+\lo,\rowa-\dl ) -| (\colb-\lo,\rowa);
\draw [line width=\lt] (\colb+\lo,\rowa) -- (\colb+\lo,\rowa-\dl ) -| (\colc-\lo,\rowa);
\draw [line width=\lt] (\colc+\lo,\rowa) -- (\colc+\lo,\rowb);
\draw [line width=\lt] (\cola+\lo,\rowb) -- (\cola+\lo,\rowb-\dl ) -| (\colc-\lo,\rowb);
\draw [line width=\lt] (\cola,\rowb) -- (\cola,\rowb+\dl+\dl/2 ) -- (\colb,\rowa-\dl ) -- (\colb,\rowa);
\draw [line width=\lt] (\cola+\lo,\rowb) -- (\cola+\lo,\rowb+\dl ) -- (\colc,\rowa-\dl-\dl/2 ) -- (\colc,\rowa);


\node [load,shape border rotate=0,scale=\ls] at (\colb,\rowa+\dld) {};
\draw (\colb,\rowa) to[short] (\colb,\rowa+\dld);
\node [load,shape border rotate=0,scale=\ls] at (\colc,\rowa+\dld) {};
\draw (\colc,\rowa) to[short] (\colc,\rowa+\dld);
\node [load,shape border rotate=180,scale=\ls] at (\colc,\rowb-\dld) {};
\draw (\colc,\rowb) to[short] (\colc,\rowb-\dld);
\node [load,shape border rotate=180,scale=\ls] at (\cola-\lo,\rowb-\dld) {};
\draw (\cola-\lo,\rowb) to[short] (\cola-\lo,\rowb-\dld);

\end{circuitikz}
\end{center}

\vspace{-7mm}
        \caption{Five bus system} 
        \label{fig_three_bus}
    \end{subfigure}
    \caption{Systems for testing Quantum Power Flow (QPF).}
    \label{fig_systems}
\end{figure}
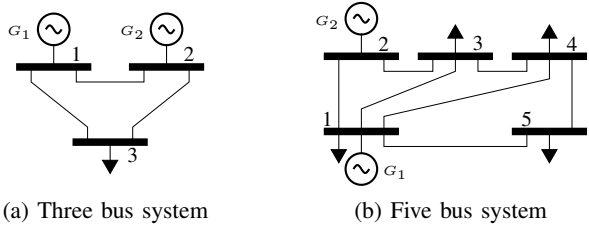

The capabilities of both real and simulated Quantum Computers are still limited. Some simplifications must therefore be made to run the power flow algorithm on a real quantum computer. 
If each of the eigenvalues of the system matrix cannot be exactly represented by the same number of classical bits as there are qubits in the $nl$ register, the quantum phase estimation will only give an approximation of the eigenvalues $\lambda_j$ of the $B'$ and $B''$ matrices; this introduces an error in the results. In order to investigate the performance of real quantum computers, at this stage, we select the parameters of the test systems so that each of the eigenvalues of the $B$ matrices can be closely represented by 3 bits. This will correspond to the need of $nl=3$ qubits in our quantum circuit. As our focus is to test a simple system on a real QC, for the purpose of the investigations in this paper we also consider only slack and PQ buses, and no phase-shifting transformers.
This means that $B^{'}=B^{''}$, and we can use the same HHL circuit for both \eqref{eq_dth2} and \eqref{eq_dv} only with a different vector $b$ loaded onto the $nb$ register (i.e., $\Delta \theta$ or $\Delta V$). The two power systems considered for this paper are shown in Fig.~\ref{fig_systems} and the parameters of the 3-bus system are shown in Table \ref{tab_params}.

\begin{table}[ht!]
\centering
\caption{Three bus system parameters}
\begin{tabular}{c|ccccc}
\textbf{Bus} & \textbf{Type} &$\mathbf{P_{MW}}$ & $\mathbf{Q_{MVAr}}$ & $\mathbf{V_{pu}}$ & $\bm{\theta^\circ}$ \\ \hline
1       & Slack     & 5         & 6.9       & 1.03      & 0     \\
2       & PQ        & 10        & 0         &  1.012    & 2.109     \\
3       & PQ        & -15       & -5        & 0.985     & -4.996      
\end{tabular}\\
\vspace{3mm}

\begin{tabular}{c|cccc}
\textbf{Line} & \textbf{From} & \textbf{To}  & $\mathbf{R_{pu}}$ & $\mathbf{X_{pu}}$\\ \hline
1                 & 1             & 2        & 0           & 1j     \\
2                 & 1             & 3        & 0           & 1j     \\
3                 & 2             & 3        & 0           & 2j      
\end{tabular}
\label{tab_params}
\end{table}
The parameters in Table~\ref{tab_params} give the $B$ matrices:
\begin{equation}
\boldsymbol{B^{'}} = \boldsymbol{B^{''}} =
\begin{bmatrix}
-1.5 & 0.5 \\
0.5 & -1.5 \\
\end{bmatrix}
\end{equation}
The eigenvalues of matrices $\boldsymbol{B^{'}}$ and $\boldsymbol{B^{''}}$ are $\{-1, -2\}$.




\subsection{Quantum hardware requirements}

The number of qubits (circuit width) required for the HHL circuit depends on the required accuracy of the phase estimation and the number of variables in the data vector (plus the auxiliary qubit that we always need to indicate if we have obtained a binary estimation or not). For example, the 3-bus system has 2 variables for $\Delta P$ which are encoded in 1 qubit ($\#qubits=log_2 N$, where $N$ is the number of variables, as mentioned in Section~\ref{sec:qc}) and we use a 2-bit estimation of the eigenvalues of $B^{'}$. Another qubit is needed for the phase estimation in case the system matrix has negative eigenvalues. Including the auxiliary qubit, this gives a total of 5 qubits required for the 3-bus system. The qubit for negative eigenvalues could be eliminated if we can assume all eigenvalues have the same sign. This is, however, not always the case for larger power systems which could for example include capacitive branches.

We shall note here that the true benefits of Quantum Computing emerge in large systems, since the number of required qubits have a sublinear increase with larger system sizes (logarithmic to be precise). This means that we can represent very large systems with only a few qubits. For example, applying the $log_{2}(N)$ rule for the required number of qubits, we can represent the data vector of a 500'000-bus system with only 20 qubits ($2^{20}=1$'$048$'$576 \approx 2\times 500$'$000$ values). 

Currently, however, the major issue with scalability is not as much the circuit width, i.e. the number of qubits, but the circuit depth; that is, the number of gates in the quantum circuit, as shown in Fig.~\ref{fig_circuit} . The current general implementation of the HHL algorithm uses methods which scale exponentially with the number of qubits, and, so far, no algorithm exists for preparing the HHL circuit with polynomially increasing resources for an arbitrary matrix. Table \ref{tab_qc_sizes} shows how the circuit sizes grow with the matrix size. 

\begin{table}[ht!]
\centering
\caption{Quantum Circuit Sizes}
\begin{tabular}{l|cccc}
\textbf{} & \textbf{3 bus} & \textbf{5 bus} & \textbf{9 bus} & \textbf{17 bus} \\ \hline
$B^{'}$ matrix size                 & 2x2             & 4x4           &   8x8 & 16x16   \\
Circuit Width                 & 5             & 7           &  9  &     11   \\
Circuit Depth                  & 336             & 3528           & 76876   &   802737     \\
CNOT Gates                  & 108             & 1181           & 28956    &     298594  
\end{tabular}
\label{tab_qc_sizes}
\end{table}

\section{Simulation results}
\label{sec:results}

\subsection{3-bus system on real Quantum Computers}
The power flow application is implemented with IBM's qiskit (0.34.1). The 3-bus system is tested on 4 of IBM's open access quantum computers \cite{ibmq} listed in Table \ref{tab_qc}: ibmq\_lima, ibmq\_belem, ibmq\_quito, and ibmq\_bogota.  The quantum computers listed in Table~\ref{tab_qc} have different sizes, configurations and error rates. The Quantum Volume (QV) is a measure of the device's performance regardless of the underlying technology or the number of qubits.

\begin{table}[ht!]
\centering
\caption{Quantum Computers used for testing Quantum Power Flow (values taken on 23.02.2022)}
\begin{tabular}{l|ccc}
\textbf{QC} & \textbf{Qubits} & \textbf{QV} & \textbf{avg. CNOT error} \\ \hline
ibmq\_lima                 & 5             & 8           &   9.996e-3  \\
ibmq\_belem                & 5             & 16           &  1.363e-2  \\
ibmq\_quito                  & 5             & 16           & 1.135e-2 \\
ibmq\_bogota                 & 5             & 32           & 1.206e-2 \\ 
ibm\_perth                  & 7             & 32            & 1.006e-2
\end{tabular}
\label{tab_qc}
\end{table}

The results of the 3-bus power flow on ibmq\_quito are shown for each iteration of the power flow calculation in Table~\ref{tab_iter}. The rest of the quantum computers we used, i.e. ibmq\_lima, ibmq\_belem, and ibmq\_bogota give similar results.
The Quantum Power Flow converges to the same solution as the classical FDLF but, due to the noisy hardware available at the moment, it requires a much larger number of iterations. With the tolerance set to $10^{-5}$, the classical method converges in 5 iterations while the Quantum Power Flow (QPF) takes around 32-38 iterations. 

The available quantum software (qiskit 0.34.1) does not allow us to implement if-loops, for-loops, and while-loops within a real quantum algorithm at the moment (March 2022). As a result, implementing a numerical solution algorithm requires us to extract the value of the quantum data vector in every iteration, perform the logical operations (e.g. is $\Delta P \geq \xi$, see Algorithm 1) and plug $\Delta P$ back in the quantum circuit again for the next iteration. Considering that the data vector is in a quantum state, we need to ''probe'' $\Delta P$ multiple times to measure the probability distribution of the solution. This requires time of at least $\mathcal{O}(N)$ (see also the Discussion in Section~\ref{sec:discussion}).  Obviously, this slows down our quantum algorithm considerably at the moment. The next qiskit version, which is expected around August 2022, is expected to allow for extended functionality; this will unleash a wide range of opportunities, including if-clauses, for-loops, and while-loops inside the quantum algorithm which could eliminate the need for measuring the data vector in each iteration of the QPF.

\begin{figure}
    \centering
    \begin{subfigure}[b]{\columnwidth}
        \includegraphics[width=\columnwidth]{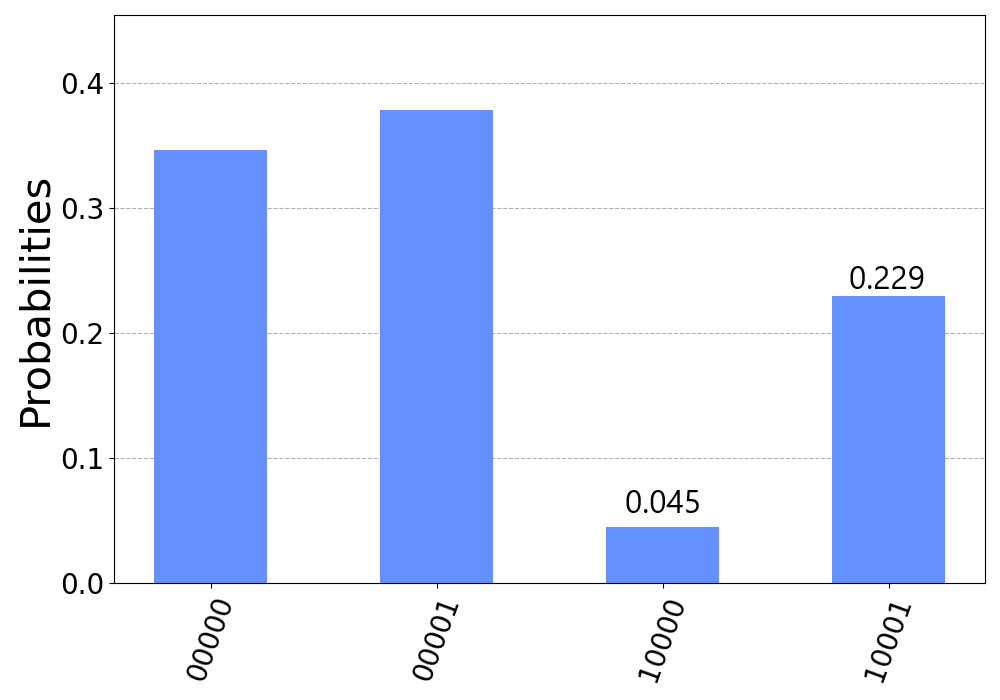}
        \caption{Without noise} 
    \end{subfigure}\\%
    \begin{subfigure}[b]{\columnwidth}
        \includegraphics[width=\columnwidth]{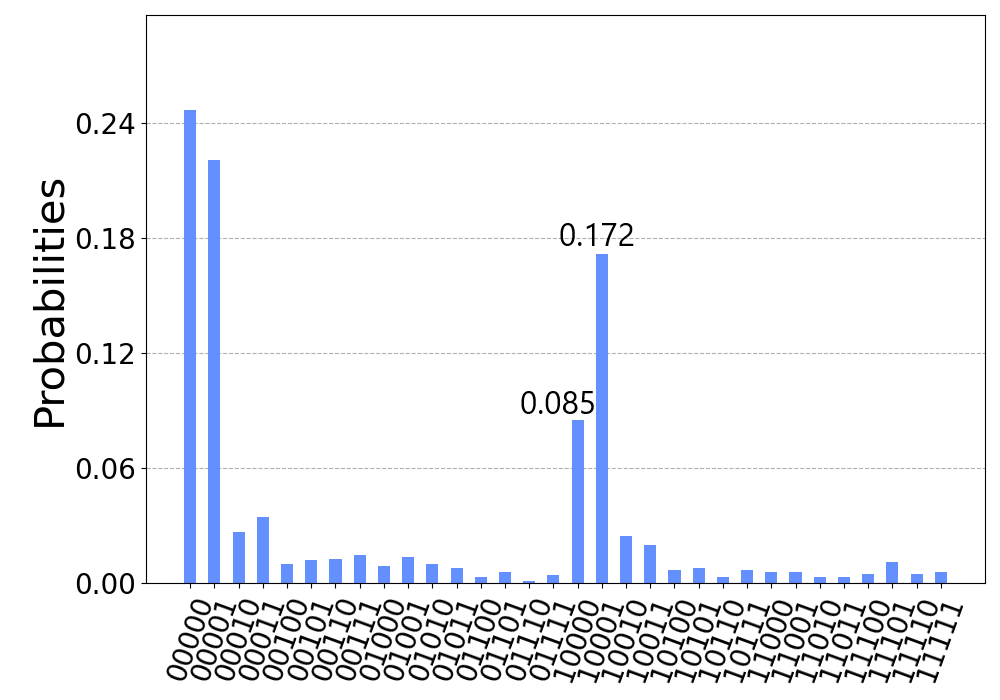}
        \caption{With noise} 
    \end{subfigure}
    \caption{Histogram showing measurements of $\ket{\Delta \theta}$ in the first iteration. The values of $\Delta \theta$ are estimated from the probability distribution of the data read in register $nb$ (see Fig.~\ref{fig_circuit}), when the auxiliary bit $na=1$ i.e. from the states  ``10000'' and ``10001''
    }
    \label{fig_fig_hist}
\end{figure}

Going back to the currently possible implementation of our QPF, where the data vector $\Delta P$ needs to be measured in every iteration, Fig.~\ref{fig_fig_hist} shows the measurement from the first iteration with and without noise. The derivation of \eqref{eq_dth2} assumes perfect phase estimation. Based on the HHL circuit, that means that qubits in the $nl$ register will be in the state $\ket{0}$ after the inverse phase estimation, and the result state is conditioned on seeing the auxiliary qubit in register $na$, as 1. In that case, the normalized values of $\Delta \theta$ can be read from the $nb$ register, i.e. the states ``10000'' and ``10001'' in the histogram in Fig.~\ref{fig_fig_hist}.  

The quantum circuit in Fig.~\ref{fig_circuit} is executed 1024 times to retrieve a probability distribution of the quantum states as shown in Fig.~\ref{fig_fig_hist}. The probabilities shown in the histogram sum up to 1 where the amplitudes of the quantum state coefficients satisfy $\left| a \right|^2+\left| b \right|^2+... = 1$. To get an estimate of the amplitudes of $\Delta \theta$, as we also explained in Section~\ref{sec:pf}, we must therefore take the square root of the probability and scale it with the norm of the input $\Delta P$.

\begin{equation}
    \Delta \theta \approx \lVert \Delta P \rVert*\sqrt{probability}
    \label{eq:qubit_prob}
\end{equation}

\begin{figure}[ht!]
\centering
\includegraphics[width=0.95\columnwidth]{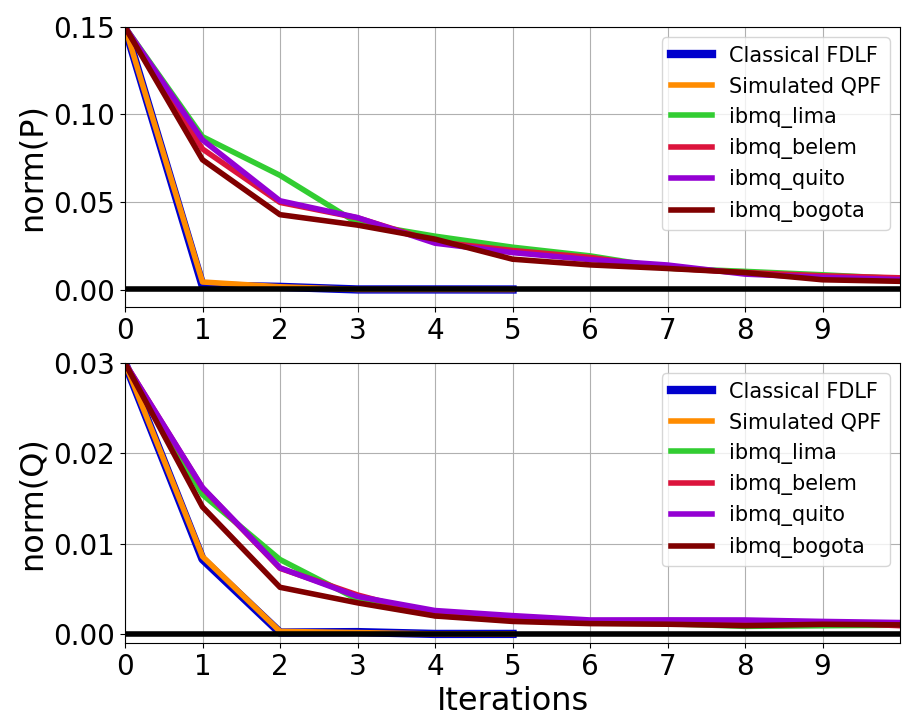}
\caption{Power flow convergence of the 3-bus system}
\label{fig_convergence}
\vspace*{-\baselineskip}
\end{figure}

For the first iteration of the noise free simulation, this gives:
\begin{equation}
\begin{bmatrix}
\Delta \theta_2 \\
\Delta \theta_3 \\
\end{bmatrix} \approx 0.1803*
\begin{bmatrix}
\sqrt{0.045} \\
\sqrt{0.229} \\
\end{bmatrix} \approx 
\begin{bmatrix}
0.03824 \\
0.08628 \\
\end{bmatrix}
\end{equation}
Where $\lVert \Delta P \rVert=0.1803$. We can see that the result of the noise-free quantum simulation is very close to the result for the first iteration of the classical computing algorithm, which is  $\Delta \theta = [0.0375,  0.0875]$ rad. On the contrary, for noisy quantum computations, which are based on real quantum computers, the output is $\Delta \theta = [0.0526, 0.0748]$ rad after the first iteration. This is the reason why noise-free quantum computing converges almost equally fast as classical computing (same number of iterations, see Fig.~\ref{fig_convergence}), while the algorithms implemented on actual quantum computers require at the moment a significantly larger number of iterations.

\begin{table}[]
    \centering
    \caption{Iterations of Classical FDLF vs QPF on ibmq\_quito}
    \begin{tabular}{>{\centering\arraybackslash}b{1.3cm}|cccc}
    \textbf{FDLF iteration} & $V_2$ & $\theta_2$ & $V_3$ & $\theta_3$ \\ \hline
        1 & 1.01750000 & 2.14859173 & 0.99250000 & -5.01338071\\
        2 & 1.01198104 & 2.08682333 & 0.98513520 & -4.94593291\\
        3 & 1.01210909 & 2.10823443 & 0.98510531 & -4.99645096\\
        4 & 1.01200181 & 2.10832465 & 0.98496362 & -4.99531212\\
        5 & 1.01200181 & 2.10872362 & 0.98496362 & -4.99629051\\
    \end{tabular}\\
    \vspace{2mm}
    \begin{tabular}{>{\centering\arraybackslash}b{1.3cm}|cccc}
    \textbf{QPF iteration}& $V_2$ & $\theta_2$ & $V_3$ & $\theta_3$ \\ \hline
       1 &  1.00602829 & 2.05419579 & 0.99405120 & -1.77164967 \\
       2 &  1.00937372 & 3.12147980  & 0.99099606 & -2.73581045\\
       3 &  1.01129425 & 2.50579482  & 0.04373437 & -3.30344432\\
       $\vdots$ & $\vdots$  & $\vdots$  & $\vdots$  \\
      34 &  1.01200263 & 2.10882674  & 0.98496970 &  -4.99558292 \\
      35 &  1.01200492 & 2.10901696  & 0.98496697 & -4.99577486
    \end{tabular}
    \label{tab_iter}
\end{table}

A comparison between the convergence of the classical method, the 4 physical QC devices and a simulated noise-free QC is shown in Fig.~\ref{fig_convergence}. It shows how a simulated noise-free QC gives close to identical results with a classical computer, while the real noisy QCs converge much slower. As the reader observes, there is no notable difference between the implementations on the 4 real QCs. 

\subsection{5-bus system}
We also test our QPF on a 5-bus system. Our goal is to investigate how current noisy hardware impact the Quantum Power Flow solution as we scale to larger systems. As shown in Table~\ref{tab_qc_sizes}, a QPF for a 5-bus system requires 7 qubits and a much larger number of gates. We therefore use the 7 qubit machine "ibm\_perth", shown in Table~\ref{tab_qc}, to test the 5 bus system. 

\begin{figure}[ht!]
\centering
\includegraphics[width=0.95\columnwidth]{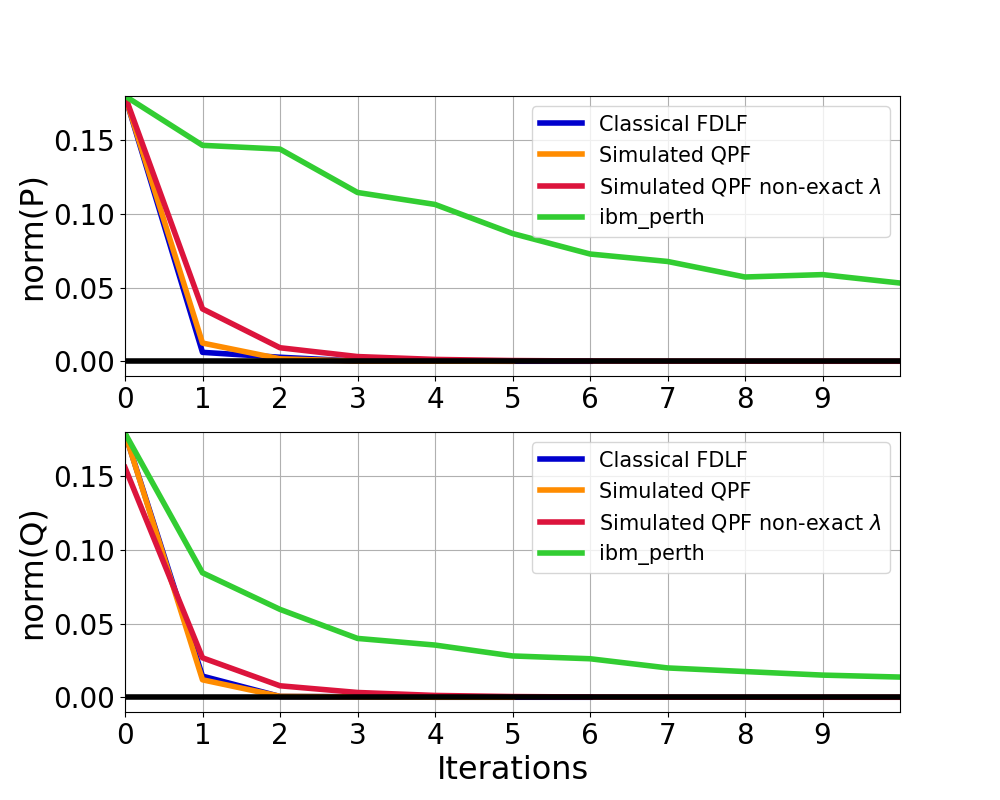}
\caption{Power flow convergence of the 5-bus system}
\label{fig_convergence5}
\vspace*{-\baselineskip}
\end{figure}

Similar to the 3-bus system, the line parameters of the 5 bus system are chosen so the eigenvalues of the $\mathbf{B}$ matrices can be closely represented by 3 bits, as our primary focus is to examine the impact of the noisy quantum hardware. The $\mathbf{B}$ matrices are presented in \eqref{eq_b5}. Fully acknowledging though that the characteristics of real power systems can vary significantly, in our experiments we also perturb the system characteristics (and the $\mathbf{B}$ matrices shown in \eqref{eq_b5}) to explore cases where $\mathbf{B}$ matrices result to eigenvalues that cannot be accurately represented by 3 bits (see Fig.~\ref{fig_convergence5}). 

\begin{equation}
\boldsymbol{B^{'}} = \boldsymbol{B^{''}} =
\begin{bmatrix}
-4 & 0.03  & 0 & 0\\
0.03 & -3  & 0.02 & 0\\
0 & 0.02& -1.55 & 0.5 \\
0 & 0 & 0.5 & -1.45 \\
\end{bmatrix}
\label{eq_b5}
\end{equation}

The eigenvalues of $\boldsymbol{B^{'}}$ and $\boldsymbol{B^{''}}$ in this case are  $\{-1,-2,-3,-4\}$.

Figure~\ref{fig_convergence5} shows the iterations of the 5-bus power flow. As with the 3-bus system, it converges to the correct solution and the noise-free simulation is very close to the classical one. When we alter the line parameters, so that the eigenvalues of the $\mathbf{B}$ matrices cannot be represented exactly by 3 bits, we observe that this introduces a small delay in the convergence (red line in Fig.~\ref{fig_convergence5}) and if the eigenvalues vary greatly then the QPF will not converge at all without using more qubits. The major difference in convergence speed emerges when we add the noise characteristics in the computation (green line in Fig.~\ref{fig_convergence5}). 

As before, the presence of noise in the real hardware greatly impacts the number of iterations required. The larger number of gates required for the circuit compared to the 3-bus system also means greater impact from the noise, as each gate adds noise to the result. The 5-bus system therefore requires much larger number of iterations to converge on real hardware. Our simulations show that unless we develop low-noise quantum hardware, it would be challenging to perform computations for larger systems.

Based on the findings from our experiments, in the next section, we discuss opportunities and challenges arising from the development of quantum algorithms for power system applications. 
\section{Discussion}
\label{sec:discussion}

The result given by the HHL algorithm is encoded as a quantum state. If we want to extract the full state vector from the quantum state we need time of at least $\mathcal{O}(N)$. Therefore, if we do this in every iteration of our power flow we lose any benefit of the quantum speedup \cite{Aaronson2015}. 
For an HHL-based AC QPF to achieve quantum advantage, the whole iteration process would need to be performed by the QC, so that we only extract the final result. To do that, we need some form of quantum memory in order to be able to perform logical operations such as if-then-else clauses, for-loops and while-loops. 
The quantum memory will allow us to store the quantum state between iterations so we can update our power mismatches and perform the logical operations (e.g. is $\Delta P \geq \xi$,
see Algorithm 1) inside the QC algorithm, without requiring us to extract e.g. the $\Delta P$ in every iteration, check if it is greater than the defined tolerance $\xi$ and plug it back to the QC algorithm. Note here that every time we have to extract a quantum state, e.g. determine the value of $\Delta P$ outside the QC algorithm, we need to ”probe” $\Delta P$ multiple times to measure the probability distribution of the solution. This requires time of at least $\mathcal{O}(N)$, which slows down our quantum algorithm considerably at the moment. The functionality that will allow the logical operations inside the QC algorithms is not yet available in current real QC machines. The next release of qiskit, probably in August 2022, is expected to include some form of looping capabilities which will greatly improve what we are capable of doing with our quantum algorithms.

Even when we have some form of quantum memory, the process of extracting the final results can still reduce the speedup of QC, as we need to probe the output register several times to extract accurately the probability distributions (see e.g. Fig.~\ref{fig_fig_hist} for an example of the extracted histogram for the $\Delta P$ output of the 3-bus system).
Still, however, it is possible to obtain some limited statistical information, such as the presence of very high values, from the HHL algorithm without extracting the full solution vector, i.e. without requiring us to probe the output register too many times. This could be useful for contingency assessment where we have a large number of scenarios and only want to identify the ones that are unstable or overloaded. More research is necessary to identify the opportunities of such approaches.

 A further discussion point that is often raised relates to the number of qubits required for larger systems. Indeed, a larger number of qubits will be required to implement a more general QPF method capable of accurately representing both real and complex eigenvalues of the system matrices of real power systems. However, the number of required quantum states scale only logaritmically with the number of variables, i.e. $\#qubits = \log_{2}(N)$, where $N$ is the number of variables.  For example, as mentioned in Section~\ref{sec:sim}, applying the $log_{2}(N)$ rule for the required number of qubits, we can represent the data vector of a 500'000-bus system with only 20 qubits ($2^{20}=1$'$048$'$576 \approx 2\times 500$'$000$ values). Same is the case for the accuracy of the eigenvalues of the system matrices. Every qubit we add increases the accuracy of the eigenvalues by one order of magnitude. With the rapid development in quantum hardware, we expect that this will not be the limiting factor for the QPF. 

\begin{figure}[h!]
\centering
\includegraphics[width=0.8\columnwidth]{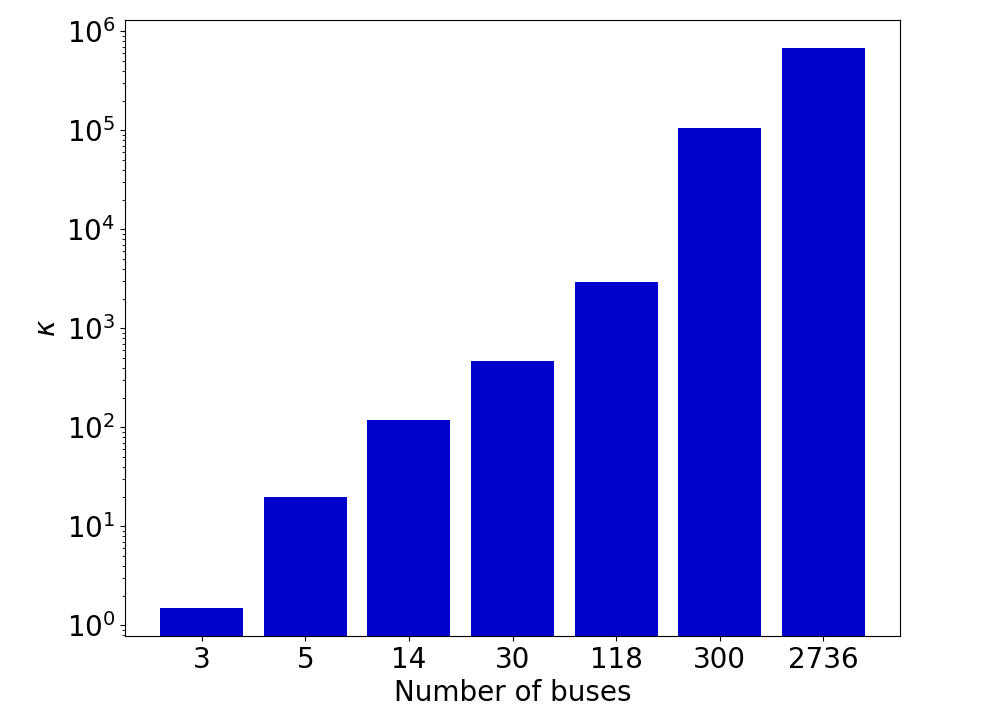}
\caption{Condition number $\kappa$ of $B^{'}$ for 6 IEEE test systems (3 - 300 buses) and the 2736-bus Polish system.}
\label{fig_condition}
\end{figure}

For the HHL algorithm to be efficient, the B matrix should be well-conditioned. As B is Hermitian, the condition number is given by the ratio between the largest and smallest eigenvalue $\kappa=\frac{|\lambda_{max}|}{|\lambda_{min}|}$. If $\kappa$ grows significantly with the size of B, then the exponential speedup of HHL is lost \cite{Aaronson2015}. As shown in Fig.~\ref{fig_condition}, the condition number seems to grow exponentially for larger test cases. This means that there is a need to design a preprocessing procedure for the system matrices to reduce their condition number while maintaining their characteristics, before plugging them in a Quantum Computer, if we wish to maintain the quantum advantage. Further research is required on how to achieve that in order to enable efficient quantum computations.

The quantum power flow converges to the correct solution even with current noisy hardware. However, the high number of iterations required could make it challenging to achieve quantum advantage while quantum computers have a high level of noise. We expect that applications of quantum computing will become attractive when the noise levels decrease. A growing number of scientists and engineers are currently working on reducing the noise level, as this will enable a wide range of new applications. This is expected to happen in the following years. Meanwhile, power system researchers can focus on inventing the methods that can adapt existing algorithms and processes to take advantage of the emerging quantum computing capabilities.

The implementation of QPF, as presented in this paper, aims to adapt the classical FDLF method to run on a quantum computer. As we showed, there is a number of challenges with current available hardware and the lack of quantum memory. However, since quantum computers are fundamentally different from classical computers, future implementations of QPF might even take a different approach in order to better utilize the unique capabilities of QC of evaluating multiple values simultaneously and possibly avoid the iterative process altogether. We identify here two key opportunities. First, the ability to drastically accelerate approaches that require some form of probability distribution, e.g. Monte-Carlo methods. The fact that quantum states can encode a probability distribution could for example significantly accelerate the computation of the mean and value-at-risk \cite{Cristina_Sanz_QMC}. This can help accelerate, for example, N-1 computations and help assess whether a probability distribution of generation profiles (due to e.g. uncertain wind power and solar PV infeed) leads to N-1 violations. We plan to investigate this in our future work. Another opportunity relates to the Power Flow computation itself, which belongs to the wider family of numerical algorithms for non-linear systems. Designing a method that can efficiently implement the equivalent of a Newton-Raphson algorithm in a Quantum Computing environment can unleash opportunities not only for power systems for a very large family of problems in analysis, optimization, and control. Intensive research efforts are required towards these directions to uncover the still unexplored potential of quantum computing.

Last but not least, the new capabilities that QC brings along require us to think differently. QC has been shown to solve problems that have so far been impossible with classical computing; such an example is the creation and observation of a new phase of matter, popularly known as time crystals \cite{Nature_time_crystals}. Pursuing an out-of-the-box thinking, where instead of replicating existing power system algorithms from the classical computing domain to the QC domain,  could uncover opportunities for power system computation, assessment, and control that have so far been impossible with classical computing. Here, we need to stress that as QC researchers say, we talk about ``quantum advantage'' and not about ``quantum supremacy''. Our goal is not to completely replace classical computing and high performance computing (HPC). On the contrary, our goal shall be to identify which are the processes that Quantum Computing can do well and deliver new capabilities, by complementing the strengths offered by HPC.


\section{Conclusion}
\label{sec:conclusion}

In this paper, we have successfully implemented and tested, for the first time to our knowledge, a quantum AC power flow application on real Noisy-Intermediate-Scale-Quantum-era (NISQ-era) quantum computers. We have shown that current hardware is capable of performing a power flow for small test systems, but scalability is currently a major issue.

There are still several challenges to be resolved before practical quantum power flow applications can achieve quantum advantage. But this also creates room for major opportunities. In our paper, we try to identify potential directions for future research that can reap the benefits of quantum computing. With further development and increasing capabilities of quantum computers, quantum applications for power systems could become extremely useful for future power system analysis, control, and optimization. An increased focus by the power system researchers on this area can lead to the development of the necessary quantum computing methods that can exploit these benefits as soon as they become available.

\section{Acknowledgements}
The authors would like to thank Jan Lillelund, CTO of IBM Denmark, and Gianni del Bimbo, Head of Quantum Engineering, and Cristina Sanz-Fernandez, Quantum Software Developer, from Multiverse Computing for the insightful discussions.

\bibliographystyle{IEEEtran}
\bibliography{refs}
%

\end{document}